\documentclass[aps, prd, amsmath, floats, floatfix, twocolumn, nofootinbib, nosuperscriptaddress]{revtex4-1}
\usepackage[pdftex]{graphicx}
\usepackage{color}
\usepackage{latexsym}
\usepackage[colorlinks]{hyperref}

\newcommand{\be}{\begin{eqnarray}}
\newcommand{\ee}{\end{eqnarray}}

\begin{document}

\title{Astrophysical Black Holes: A Review}

\author{Cosimo~Bambi}
\email{bambi@fudan.edu.cn}
\affiliation{Center for Field Theory and Particle Physics and Department of Physics, Fudan University, 200438 Shanghai, China}

\begin{abstract}
In this review, I have tried to focus on the development of the field, from the first speculations to the current lines of research. According to Einstein's theory of general relativity, black holes are relatively simple objects and completely characterized by their mass, spin angular momentum, and electric charge, but the latter can be ignored in the case of astrophysical macroscopic objects. Search for black holes in the sky started in the early 1970s with the dynamical measurement of the mass of the compact object in Cygnus~X-1. In the past 10-15~years, astronomers have developed some techniques for measuring the black hole spins. Recently, we have started using astrophysical black holes for testing fundamental physics. 
\end{abstract}

\maketitle


\section{Introduction}

\subsection{Historical overview}

Roughly speaking, a black hole is a region of the spacetime in which gravity is so strong that nothing can escape or communicate with the exterior region. Speculations on the existence of similar systems can be dated back to the end of the XVIII century. John Michell in 1783 and, independently, Pierre-Simon Laplace in 1796 discussed the possibility of the existence of extremely compact objects, so compact that the escape velocity on their surface could exceed the speed of light. For a spherically symmetric body of mass $M$ and in the framework of Newtonian mechanics, one finds that the escape velocity exceeds the speed of light $c$ if the radius of the body is smaller that 
\be
R = \frac{2 G_{\rm N} M}{c^2} \, .
\ee
Such hypothetical objects were called {\it dark stars} because they had to appear dark being unable to emit radiation from their surface. Apparently the proposal of Michell and Laplace remained at the level of a theoretical speculation and nobody tried to look for similar objects in the sky.

At the end of 1915, Albert Einstein proposed the theory of general relativity~\cite{einstein}. The simplest black hole solution was discovered immediately after, in 1916, by Karl Schwarzschild and is now called the {\it Schwarzschild solution}~\cite{schwarzschild}. It describes the spacetime of a spherically symmetric and electrically uncharged black hole. The solution describing a spherically symmetric black hole with a possible non-vanishing electric charge was found by Hans Reissner in 1916 and Gunnar Nordstr\"om in 1918 and is now called the {\it Reissner-Nordstr\"om solution}~\cite{reissner,nordstrom}. Note, however, that the actual nature of these solutions was not understood for a while. Schwarzschild was simply looking for the spherically symmetric vacuum solution of Einstein's equations describing the exterior gravitational field of a spherically symmetric and electrically uncharged body, like it is approximately that of a star or a planet. Reissner solved Einstein's equations for a point-like charged mass and Nordstr\"om solved the equations for a spherically symmetric charged mass. These solutions were singular on a surface, which is now known to be the location of the event horizon, but the nature and the implications of such a surface were not understood. In was only in 1958 that David Finkelstein realized the actual nature of these solutions and described the event horizon as a one-way membrane, so that whatever crosses the event horizon cannot influence the exterior region any longer~\cite{finkelstein}.

The solution describing the spacetime of a rotating and electrically uncharged black hole was found in 1963 by Roy Kerr and is now called the {\it Kerr solution}~\cite{kerr}. This was a very important discovery for astrophysical applications, because astronomical bodies have naturally a non-vanishing spin angular momentum. The complete solution describing a rotating and charged black hole was found in 1965 by Ezra Newman and collaborators and is now called the {\it Kerr-Newman solution}~\cite{newman}.

Since the late 1960s, people realized that the black holes of general relativity are simple objects, in the sense that they are completely specified by a small number of parameters~\cite{israel,carter,robinson}: their mass $M$, their spin angular momentum $J$, and their electric charge $Q$. There are certain assumptions behind and the result goes under the name of {\it no-hair theorem}, but actually it is a family of theorems because there are different versions and different extensions. The name no-hair comes from the fact that people are mainly characterized by their hairs (short/long, straight/curly, black/brown/blond, etc.). Black holes are characterized by $M$, $J$, and $Q$, so these are their hairs, but since these are just three parameters it follows that black holes have almost no hairs.

The Schwarzschild solution describes the exterior gravitational field of a spherically symmetric body like, with a good approximation, a star or a planet. The singular surface in the Schwarzschild solution was initially thought to have no physical implications because its radial coordinate is much smaller than the physical radius of any typical astronomical body, where the exterior solution should be pasted with the interior non-vacuum solution describing the gravitational field inside the object. In the 1920s, it was known that when a star finishes its nuclear fuel it shrinks to find a new equilibrium configuration, and that at a certain point the quantum pressure of electrons stops the collapse and the object becomes a white dwarf. In 1931, Subrahmanyan Chandrasekhar showed that there is a critical mass, now called the {\it Chandrasekhar mass}, above which the quantum pressure of electrons cannot stop the process and the whole body had to collapse to a point~\cite{chandrasekhar}. Such a scenario was criticized by many physicists, arguing the existence of a yet unknown mechanism capable of stopping the collapse. It was then found that a dead star with a mass exceeding the Chandrasekhar limit had to transform into a neutron star and that the quantum pressure of neutrons could stop the collapse. In 1939, Robert Oppenheimer and George Volkoff found that even neutron stars have a maximum mass and that the quantum pressure of neutrons cannot stop the collapse if the mass of the body exceeds this limit~\cite{oppenheimer}. Once again, it was advocated the existence of a yet unknown mechanism capable of stopping the collapse.

Quasars were discovered in the 1950s and their nature was unknown for a while. In 1964, Yakov Zel'dovich and, independently, Edwin Salpeter proposed that quasars were powered by the accretion material orbiting around supermassive black holes~\cite{zeldovich,salpeter}. However, their idea was not taken very seriously at the beginning. Other proposals, like the possibility that these sources were supermassive stars, were initially considered more promising.

Cygnus~X-1 is one of the brightest X-ray sources in the sky and was discovered in~1964~\cite{bowyer}. In 1971, Thomas Bolton and, independently, Louise Webster and Paul Murdin found that Cygnus~X-1 had a massive stellar companion~\cite{bolton,webster}. From the study of the orbital motion of the companion star, it was possible to estimate the mass of the compact object. The latter exceeded the maximum value for the mass of a neutron star, and therefore the compact object in Cygnus~X-1 was identified as the first stellar-mass black hole candidate. Such a finding is a milestone in black hole astrophysics and helped to convince the astronomy community about the existence of black holes in the Universe. Since then, we have obtained an increasing number of astronomical data pointing out the existence of stellar-mass black holes in some X-ray binary systems~\cite{remillard} and supermassive black holes at the center of many galaxies~\cite{kormendy}.

For astrophysical black holes, the electric charge should be completely negligible and therefore these objects should be completely characterized only by their mass and spin angular momentum. The mass is relatively easy to measure, by studying the orbital motion of nearby stars or gas, as it was done with Cygnus~X-1 and later with other stellar-mass and supermassive black holes. In the past 10-15~years, there have been significant efforts to measure black hole spins with different methods. The two leading techniques are currently the so-called continuum-fitting method~\cite{zhang,mcclintock} and X-ray reflection spectroscopy (or iron line method)~\cite{brenneman,reynolds}. More recently, there have been efforts to use astrophysical black holes to test fundamental physics, in particular Einstein's theory of general relativity in the strong field regime~\cite{bambi,yagi,cardoso}. Thanks to significant technological progress and new observational facilities, the LIGO experiment detected for the first time the gravitational waves from the coalescence of two black holes in~2015~\cite{LIGO} and the Event Horizon Telescope collaboration released the first ``image'' of a black hole in~2019~\cite{EHT}.

It is curious that the term ``black hole'' is relatively recent and its origin is unknown. We do not know who used the term first. In 1964, the journalist Ann Ewing was the first to use this term in a publication. It was on a report on Science News Letter. The term became quickly very popular after it was used by John Wheeler at a lecture in New York in 1967.

\subsection{Physical properties}

Astrophysical black holes should be well described by the ``ideal'' Kerr solution of general relativity. When a collapsing object enters inside its own event horizon, the spacetime metric quickly reduces to the Kerr solution through the emission of gravitational waves~\cite{price}. The impact of the gravitational field of the accretion disk or of nearby stars is normally completely negligible near the black hole~\cite{bambi2,bambi3}. Any initial non-vanishing electric charge can be almost neutralized very quickly because of the highly ionized host environment around black holes and the residual equilibrium electric charge is extremely small, and thus completely negligible, for macroscopic objects~\cite{bambi4}.

The mass of the black hole, $M$, sets the size of the system. The gravitational radius is defined as
\be
r_{\rm g} = \frac{G_{\rm N} M}{c^2} = 14.77 \left(\frac{M}{10 \,\, M_\odot}\right) \text{km} \, .
\ee
The characteristic timescale is thus
\be
\tau = \frac{r_{\rm g}}{c} = 49.23 \left(\frac{M}{10 \,\, M_\odot}\right) \text{$\mu$s} \, .
\ee
For a supermassive black hole with a mass $M \sim 10^6$~$M_\odot$, we find $r_{\rm g} \sim 10^6$~km and $\tau \sim 5$~s. If the supermassive black hole has a mass $M \sim 10^9$~$M_\odot$, we find $r_{\rm g} \sim 10^9$~km and $\tau \sim 1$~hr. Variability in the electromagnetic and gravitational wave spectra associated to modifications in the configuration of the system in the strong gravity region follows these timescales.

The spin of the black hole, $J$, only affects the strong gravity region. Indeed in Newtonian gravity the spin of a body has no gravitational effects, and only its mass appears in Newton's Law of Universal Gravitation. This is not true in general relativity, where a rotating mass has some analogies with a rotating charge in electrodynamics. For example, the spin changes the position of the event horizon. In Boyer-Lindquist coordinates, the radius of the event horizon of a Kerr black hole is
\be\label{eq-h}
r_{\rm H} = r_{\rm g} \left( 1 + \sqrt{1 - a_*^2}\right) \, ,
\ee
where $a_*$ is the dimensionless spin parameter defined as
\be
a_* = \frac{c J}{G_{\rm N} M} \, .
\ee
As we can see from Eq.~(\ref{eq-h}), the spin parameter is subject to the constraint $| a_* | \le 1$. A non-rotating (Schwarzschild) black hole has spin parameter $a_* = 0$ and a maximally rotating black hole has spin parameter $a_* = \pm 1$. For $| a_*| > 1$, there is no event horizon and the Kerr metric describes the spacetime of a naked singularity. In astrophysical contexts, it is common to consider only the case $| a_* | \le 1$ and ignore the possibility of naked singularities.

\begin{figure}[t]
\begin{center}
\includegraphics[type=pdf,ext=.pdf,read=.pdf,width=8.7cm]{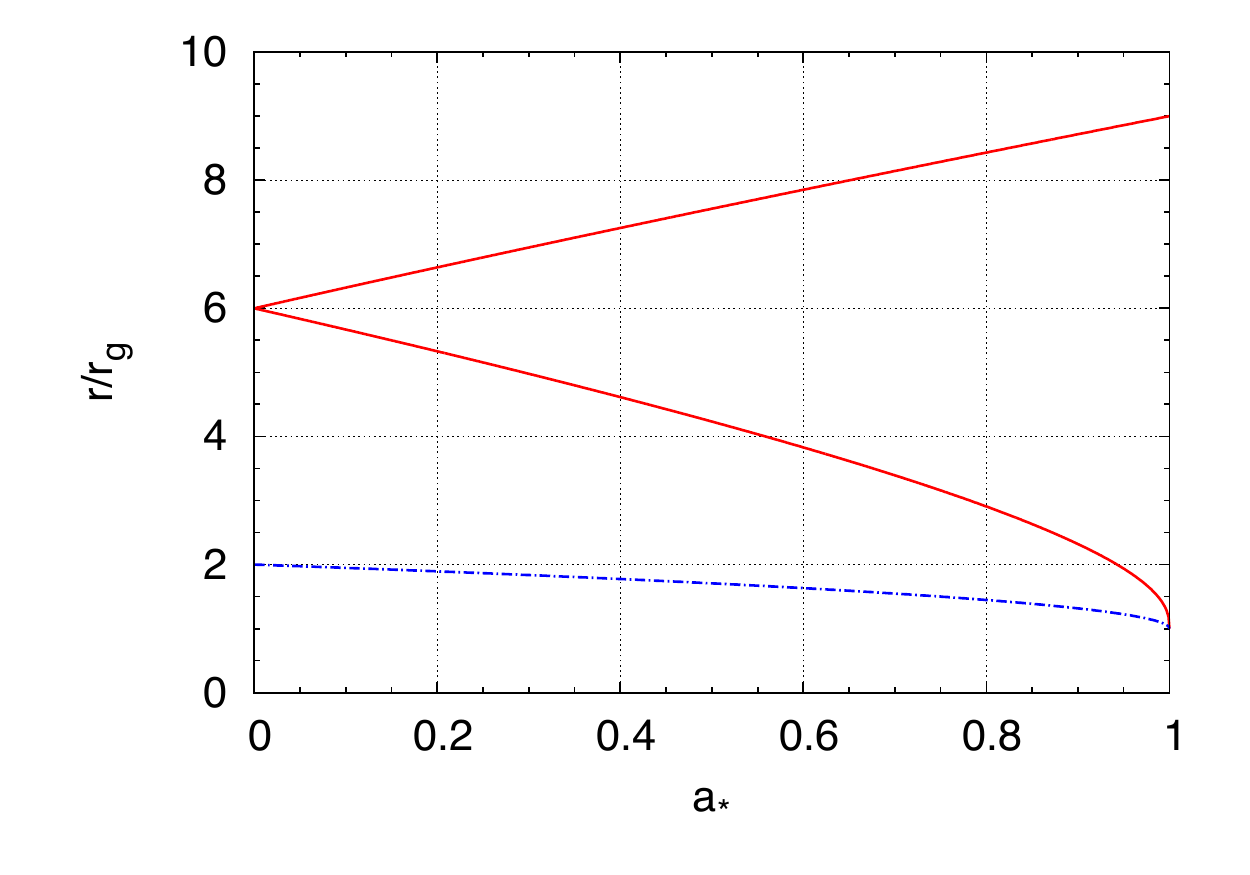}
\end{center}
\vspace{-0.6cm}
\caption{Radius of the ISCO (red solid line) and of the event horizon (blue dashed-dotted line) of a Kerr black hole in Boyer-Lindquist coordinates as a function of the dimensionless spin parameter $a_*$. For the ISCO radius, the upper curve refers to counterrotating orbits and the lower curve to corotating orbits. \label{f-isco}}
\end{figure}

The Novikov-Thorne model is the standard framework for the description of thin accretion disks around black holes and is largely employed to interpret astronomical data~\cite{ntmod}. In this model, the disk is assumed to be on the equatorial plane perpendicular to the black hole spin. The particle gas follows nearly-geodesic circular orbits in the equatorial plane. Interestingly, these orbits are not always stable but there is the so-called {\it innermost stable circular orbit} (or ISCO), $r_{\rm ISCO}$. The orbits are stable for $r > r_{\rm ISCO}$ and unstable (or they do not exist at all) for $r < r_{\rm ISCO}$. Heuristically, we can say that near a black hole the gravitational force is so strong that stable orbits are not possible. This implies that the inner edge of a thin accretion disk is at the ISCO or at a larger radius, and such a property can be exploited to measure black hole spins (see Section~\ref{s-spin})\footnote{Once the particles reach the ISCO, they quickly plunge onto the black hole.}. Fig.~\ref{f-isco} shows the radial coordinate of the ISCO radius and of the event horizon in Boyer-Lindquist coordinates. For the ISCO radius, we have two curves because the lower one refers to corotating orbits (orbits with angular momentum parallel to the black hole spin) and the upper one to counterrotating orbits (orbits with angular momentum antiparallel to the black hole spin). Heuristically, the spin has the effect to reduce the strength of the gravitational field and, in this way, corotating stable orbits can get closer to the black hole, while for counterrotating orbits we have the opposite effect and the ISCO radius increases as the spin parameter increases.

In the interpretation of astronomical observations of accreting black holes, a relevant quantity is the accretion luminosity of the source, which is convenient to measure in Eddington units. The Eddington luminosity is the maximum luminosity of a generic spherically symmetric astronomical body and is found when the pressure of the radiation luminosity balances the gravitational force towards the object. Assuming that the emitting material is a ionized gas of protons and electrons, the Eddington luminosity of an object of mass $M$ is
\be
L_{\rm Edd} &=& \frac{4 \pi G_{\rm N} M m_p c}{\sigma_{\rm Th}}  \nonumber\\
&=& 1.26 \cdot 10^{39} \left(\frac{M}{10 \,\, M_\odot}\right) \text{erg/s} \, ,
\ee
where $\sigma_{\rm Th}$ is the electron Thomson cross section and $m_p$ is the proton mass.


\section{Search for Black Holes}

The search for black hole candidates in the Universe started in the 1970s. Today we have a body of observational evidence for the existence of at least two classes of objects: stellar-mass black holes ($M \approx 3 - 100$~$M_\odot$) and supermassive black holes ($M \sim 10^5 - 10^{10}$~$M_\odot$). There is probably a third class of objects, intermediate mass black holes, with a mass filling the gap between the stellar-mass and the supermassive ones, but their nature is still controversial and they do not have yet robust mass measurements.

A comment is in order here. There is no ``conclusive'' proof that the objects that we call black holes are really black holes. However, generally speaking, this is because physics is an experimental science. We can only disprove a model/theory. We cannot confirm a model/theory, but only get some bounds. For the moment, all observational data are consistent with the idea that these objects are black holes and therefore we call them black holes. If at some point we will find that these objects have some properties not belonging to black holes, we may change idea and name, but for the moment we are in the opposite case, and all recent data including gravitational waves and black hole imaging are confirming what we expected from the theory.

\subsection{Stellar-mass black holes}

Stellar-mass black holes are the natural product of the complete gravitational collapse of heavy stars. When a star exhausts all its nuclear fuel, the gas pressure cannot balance the gravitational force any longer, and the body collapses. A significant faction of material is usually expelled away by this violent process. If the mass of the collapsing core exceeds $2-3$~$M_\odot$, there is no known mechanism capable of stopping the collapse and we have the creation of a black hole. The minimum mass of a stellar-mass black hole formed from the collapse of a progenitor star should thus be $2-3$~$M_\odot$ (the exact limit depends on the unknown matter equation of state at high densities and on the rotation of the body), because for lower masses the quantum pressure of electrons or neutrons can stop the collapse, leading to the formation of, respectively, a white dwarf or a neutron star~\cite{mass-bh1,mass-bh2}. It is currently unclear whether there is a mass gap between the populations of stellar-mass black holes and neutron stars, because observations may suggest such a gap while most theoretical models do not predict it~\cite{mass-gap}.

The maximum mass for a stellar-mass black hole is expected to be around 100~$M_\odot$~\cite{m1,m2,m3,m4}. Such an upper bound crucially depends on the metallicity of the progenitor star, namely the fraction of mass of the star made of elements heavier than helium. The mass of the black hole remnant formed from the collapse of a star depends indeed on the mass loss rate by stellar wind during the collapse, which increases with the metallicity because heavier elements have larger cross section and thus evaporate faster. In the case of low metallicity stars, some models predict a mass gap in the remnant population, and the black hole mass can either be $M < 50$~$M_\odot$ or $M > 150$~$M_\odot$. As the metallicity of the progenitor star increases, remnants with $M > 150$~$M_\odot$ disappear because of the higher mass loss rate during the black hole formation. However, other models do not predict such a mass gap, finding that very heavy stars undergo a runaway nuclear explosion that completely destroys the body, without leaving any black hole.

From stellar evolution studies, we expect a population of $10^8 - 10^9$~stellar-mass black holes in the Milky Way as the final product of the evolution of heavy stars~\cite{bhnum}. Since the discovery in the early 1970s of Cygnus~X-1, the search for stellar-mass black holes has mainly focused on the identification of compact objects in X-ray binaries with masses exceeding the maximum mass of neutron stars. In these systems, the X-ray radiation originates from the inner part of the accretion disk around the compact object. From the study of the orbital motion of the companion star, we may measure the mass function~\cite{casares}
\be
f (M) = \frac{K_{\rm c}^3 P_{\rm orb}}{2 \pi G_{\rm N}} = \frac{M \sin^3i}{\left( 1 + q \right)^2} \, ,
\ee
where $K_{\rm c} = v_{\rm c} \sin i$, $v_{\rm c}$ is the velocity of the companion star, $i$ is the inclination angle of the orbital plane with respect to the line of sight of the observer, $P_{\rm orb}$ is the orbital period of the companion star, $q = M_{\rm c}/M$, $M_{\rm c}$ is the mass of the companion star, and $M$ is the mass of the compact object. In general, it is necessary to have independent estimates of $M_{\rm c}$ and $i$ to infer the mass of the compact object $M$. If the latter exceeds the maximum mass for a neutron star, the compact object is classified as a black hole and we say that it has a dynamical measurement of the mass. Note that $M > f(M)$, and therefore if the mass function $f(M)$ exceeds the maximum mass for a neutron star we can conclude that $M$ does it too even if we do not have estimates of $M_{\rm c}$ and $i$.

\begin{figure}[t]
\begin{center}
\includegraphics[type=pdf,ext=.pdf,read=.pdf,width=8.7cm]{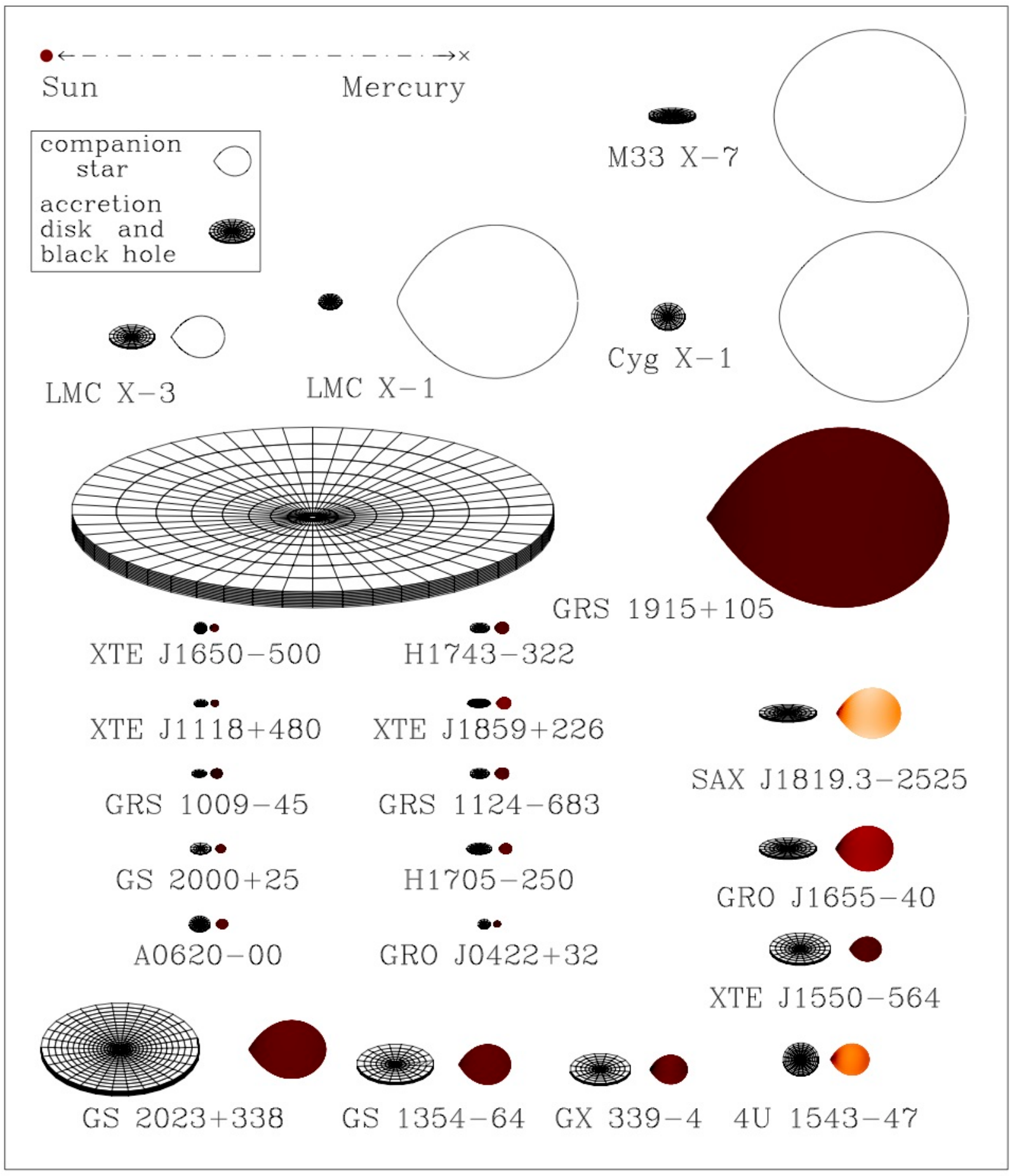}
\end{center}
\vspace{-0.3cm}
\caption{Sketch of 22~X-ray binaries with a dynamically confirmed stellar-mass black hole. For every system, we see the accretion disk around the black hole on the left and the stellar companion on the right. The color of the stellar companion refers to the surface temperature (from brown to white as the temperature increases). The orientation of the accretion disks reflects the inclination angles of the binaries. Note the system Sun-Mercury in the top left corner of the figure: the distance Sun-Mercury is about 50~millions km and the radius of the Sun is about 0.7~millions km. Figure courtesy of Jerome Orosz. \label{f-bhb}}
\end{figure}

Currently we know about 20~stellar-mass black holes in the Milky Way and a few more stellar-mass black holes in nearby galaxies. Fig.~\ref{f-bhb} shows 22~X-ray binaries with a stellar-mass black hole confirmed by dynamical measurements. Among these systems, Cygnus~X-1 (Cyg~X-1 in the figure), LMC~X-1, LMC~X-3, and M33~X-7 are high-mass X-ray binaries, in the sense that the companion star is heavy ($M > 10$~$M_\odot$). These 4~systems are persistent X-ray sources, namely they are bright X-ray sources in the sky at any time: this is because the mass transfer from the companion star to the black hole originates from the wind of the former and is a relatively stable process. The other 18~binary systems in Fig.~\ref{f-bhb} are low-mass X-ray binaries, i.e. the companion star is not heavy ($M < 3$~$M_\odot$), and they are transient sources: the mass transfer is via Roche Lobe overflow and is not continuos, so the system may be a bright X-ray source for a period ranging from a few days to a few months and then be in a quiescent state for years or decades. GRS1915+105 is quite a peculiar case because it is a low-mass X-ray binary but it is a persistent X-ray source since 1992, so for more than 25~years, probably because of the very large accretion disk that can provide material to the black hole at any time. The 22~X-ray binaries in Fig.~\ref{f-bhb} are all in the Milky Way with the exception of LMC~X-1 and LMC~X-3, residing in the Large Magellanic Cloud, and M33~X-7, residing in the nearby small galaxy M33.

Currently we also know more than 50~X-ray binaries with a stellar-mass black hole candidate, namely with a compact object that is thought to be a black hole but for which a dynamical measurement of its mass is lacking. The spectra of these sources suggest that the compact object is a stellar-mass black hole, but it is possible that in some sources there is no black hole but a neutron star. Every year the number of X-ray binaries with a stellar-mass black hole candidate increases by one or two, because we observe new X-ray sources that they were previously in a quiescent state and start emitting radiation due to the transfer of material from the companion star to the compact object.

In September~2015, the LIGO experiment detected for the first time the gravitational wave signal from the coalescence of two stellar-mass black holes, opening a completely new window to the search for these objects in the sky~\cite{LIGO}. Ground-based gravitational wave detectors now promise to discover a large number of stellar-mass black holes in the next few years. While the coalesce of two stellar-mass black holes or of a stellar-mass black hole with a neutron star can be quite a rare event in a single galaxy, current gravitational wave antennas have reached the necessary sensitivity to monitor many galaxies, making the detection of this kind of event relatively frequent, at the level of one event every few days.

\subsection{Supermassive black holes}

Astronomical observations show that at the center of every large and middle-size galaxy there is a large amount of mass in a relatively small volume~\cite{kormendy}, while in the case of small galaxies the situation is more uncertain: some small galaxies probably do have too, while other small galaxies do not~\cite{small1,small2}. In the case of the center of the Milky Way, it is possible to study the orbits of individual stars and conclude that the central mass is $M = 4 \cdot 10^6$~$M_\odot$ within a radius $R < 0.01$~pc, which is the smallest distance of the orbit of one of those stars from the center~\cite{mass-bmw1,mass-bmw2}. Such a large amount of mass in such a volume cannot be explained with a cluster of neutron stars or brown dwarfs, and the natural interpretation is that there is a supermassive black hole~\cite{maoz}. Similar conclusions can be inferred by studying the gas orbiting the center of the galaxy NGC~4258. For other galaxies, it is not possible to obtain sufficiently stringent measurements to exclude other possibilities like the presence of a cluster of neutron stars, but it is thought that all these massive systems are supermassive black holes with $M \sim 10^5 - 10^{10}$~$M_\odot$.

The recent image of the center of the galaxy M87 obtained by the Event Horizon Telescope collaboration confirms the presence of a supermassive black hole (even if some exotic alternatives cannot be ruled out)~\cite{EHT}. In particular, the dark image commonly called the shadow (even if the name may be misleading, because it is not created as a normal shadow by a body stopping light rays) is consistent with the predictions for a black hole in general relativity.

While stellar-mass black holes in the Universe are expected as the final product of heavy stars, the exact origin of the supermassive black holes at the center of galaxies is not completely understood. Heavier objects naturally tend to move to the center of a multi-body system, and it is easy to imagine that an original black hole can grow by swallowing the material around. However, it is puzzling to see objects with masses $M \sim 10^{10}$~$M_\odot$ in very distant galaxies, when the Universe was only 1~Gyr old~\cite{wu15}. It is unclear how it was possible for such black holes to grow so fast in a relatively short time. There are a few possibilities under investigation~\cite{mv}. For example, these objects may have formed from the collapse of heavy primordial clouds; if so, their initial mass was a few order of magnitude above the mass of stellar-mass black holes. They may have formed from the merger of several black holes. They may have grown faster by experiencing some period of super-Eddington accretion, which is possible within some particular accretion models~\cite{supere}. At the moment, we do not know which of these mechanisms is the correct one, nor if two or more mechanisms contributed together.

\subsection{Intermediate mass black holes}

Intermediate mass black holes would be black holes with a mass in the range $\sim 10^2 - 10^5$~$M_\odot$; that is, filling the gap between the stellar-mass and the supermassive ones~\cite{miller03}. However, the nature of these objects is still controversial. It is likely the some candidates are really intermediate mass black holes and other candidates are not.

Some intermediate mass black hole candidates are associated to the so-called ultra-luminous X-ray sources~\cite{ulxs}. Their X-ray luminosity is $L_X > 10^{40}$~erg/s and thus exceeds the Eddington luminosity of a stellar-mass object. However, without a dynamical measurement of their mass, we cannot exclude that they are neutron stars or stellar-mass black holes accreting at a moderate super-Eddington rate and with a non-isotropic emission~\cite{supere2,supere3}.

Quasi-periodic oscillations (or QPOs) are narrow features in the power density spectrum of a source. While their exact origin is currently unknown, they are associated to the short timescale variability of the source and their frequency scales as $1/M$ from stellar-mass to supermassive black holes, which is exactly what one should expect in the case of a common origin of these features for all these sources. Observations of QPOs in some ultra-luminous X-ray sources at a few Hz suggest the presence of objects heavier than normal stellar-mass black holes and lighter than normal supermassive black holes~\cite{interqpo}.

Intermediate mass black holes may be expected at the center of dense stellar clusters as the result of collisions and mergers. Studies on the dynamics of multi-body systems show that the presence of an intermediate mass black hole at the center of a cluster should increase the stellar velocity dispersion. Some observations are consistent with the presence of intermediate mass black holes in some stellar clusters, but other explanations cannot be completely ruled out~\cite{gclu1,gclu2}.


\section{Spin measurements \label{s-spin}}

In the absence of new physics, astrophysical black holes should be described by the Kerr solution and thus be completely characterized by two parameters; that is, the mass $M$ and the spin angular momentum $J$. A direct or indirect mass measurement is often the key-quantity that allows us to classify a compact object as a black hole (or black hole candidate in the case of indirect mass measurement), as we have discussed in the previous section. The spin measurement is more challenging because the spin has no effect in Newtonian gravity and thus its estimate requires the study of intrinsic relativistic effects occurring in the strong gravity region near the compact object. The development of techniques aiming at measuring black hole spins started 10-15~years ago. Currently, the two leading techniques are the so-called continuum-fitting method (suitable for stellar-mass black holes only)~\cite{zhang,mcclintock} and X-ray reflection spectroscopy (applicable to both stellar-mass and supermassive black holes)~\cite{brenneman,reynolds}. Future space-based gravitational wave antennas promise to provide very accurate measurements of the spin of supermassive black holes from the detection of extreme mass ratio inspirals (or EMRIs)~\cite{EMRIspin1,EMRIspin2}.

\subsection{Continuum-fitting method}

Geometrically thin and optically thick accretion disks around black holes have a blackbody-like spectrum locally, which becomes a multi-temperature blackbody spectrum when integrated radially. Employing the Novikov-Thorne model~\cite{ntmod}, we assume that the disk is on the equatorial plane perpendicular to the black hole spin. The gas in the disk moves on nearly geodesic circular orbits and the inner edge of the disk is set at the ISCO radius. Eventually the thermal spectrum of the disk depends on five parameters: the black hole mass $M$, the mass accretion rate $\dot{M}$, the black hole distance $D$, the inclination angle of the disk with respect to the line of sight of the distant observer $i$, and the black hole spin parameter $a_*$~\cite{kerrbb}. Since the shape of the spectrum is quite simple, it is not possible to determine all these parameters by fitting the observational data of a source. Independent measurements of $M$, $D$, and $i$ are required, and if they are available it is possible to fit the thermal spectrum and infer the black hole spin parameter $a_*$ and the mass accretion rate $\dot{M}$. This is the continuum-fitting method~\cite{zhang,mcclintock}.

This technique is normally used for stellar-mass black holes only, because the disk temperature depends on the black hole mass. For a black hole accreting at 10\% of its Eddington limit, the thermal spectrum of the disk is peaked in the soft X-ray band ($0.1-1$~keV) for stellar-mass black holes and in the optical/UV band ($1-100$~eV) for supermassive black holes. In the latter case, dust absorption prevents an accurate measurement of the spectrum and, in turn, an estimate of the spin\footnote{Despite that, there are efforts to apply the continuum-fitting method even to measure the spin of supermassive black holes~\cite{cfm-agn1,cfm-agn2}.}.

In the past 10-15~years, the spin parameter of about 10~stellar-mass black holes have been estimated with this method. It is important to select sources with a strong thermal component and accreting at 5\% to 30\% of their Eddington limit~\cite{cfm-5-30-ed}. For lower mass accretion rates, the disk may be truncated at a radius larger than the ISCO. For higher mass accretion rates, the gas pressure becomes important: the disk is not thin any longer and its inner edge may be inside the ISCO. Tab.~\ref{t-spin-stellar} shows a summary of current spin measurements of stellar-mass black holes.

\begin{table*}[t]
\centering
{\renewcommand{\arraystretch}{1.2}
\begin{tabular}{|ccccccc|}
\hline 
BH Binary & \hspace{0.1cm} & $a_*$ (Continuum) & \hspace{0.1cm} & $a_*$ (Iron) & \hspace{0.1cm} & \hspace{0.3cm} Principal References \hspace{0.3cm} \\
\hline 
GRS~1915+105 && $> 0.98$ && $0.98 \pm 0.01$ && \cite{r-bh-cfm-1915,r-bh-cfm-1915b} \\
Cygnus~X-1 && $> 0.98$ && $> 0.95$ && \cite{r-bh-cfm-cyg1,r-bh-cfm-cyg2,r-bh-cfm-cyg3,r-bh-cfm-cyg4,r-bh-cfm-cyg5,r-bh-cfm-cyg6} \\
GS~1354-645 && -- && $> 0.98$ && \cite{r-bh-cfm-gs1354} \\
MAXI~J1535-571 && --- && $> 0.98$ && \cite{r-bh-iron-maxi15a,r-bh-iron-maxi15b} \\
Swift~J1658.2 && --- && $> 0.96$ && \cite{r-bh-iron-swift-xu} \\
LMC~X-1 && $0.92 \pm 0.06$ && $0.97^{+0.02}_{-0.25}$ && \cite{r-bh-cfm-lmcx1,r-bh-cfm-lmcx1b} \\
GX~339-4 && $< 0.9$ && $0.95\pm0.03$ && \cite{r-bh-cfm-gx339,r-bh-cfm-gx339b,r-bh-cfm-gx339c,r-bh-cfm-gx339d} \\
XTE~J1752-223 && --- && $0.92 \pm 0.06$ && \cite{r-bh-cfm-1752} \\
MAXI~J1836-194 && --- && $0.88 \pm 0.03$ && \cite{r-bh-cfm-maxi} \\
M33~X-7 && $0.84 \pm 0.05$ && --- && \cite{r-bh-cfm-liu08} \\
4U~1543-47 && $0.80 \pm 0.10^\star$ && --- && \cite{r-bh-cfm-sh06} \\
IC10~X-1     &&  $\gtrsim0.7$  && --- && \cite{r-bh-cfm-st16} \\
Swift~J1753.5 && --- && $0.76^{+0.11}_{-0.15}$ && \cite{r-bh-cfm-swift} \\
XTE~J1650-500 && --- && $0.84 \sim 0.98$ && \cite{r-bh-cfm-1650} \\
GRO~J1655-40 && $0.70 \pm 0.10^\star$ && $> 0.9$ && \cite{r-bh-cfm-sh06,r-bh-cfm-swift} \\
GS~1124-683 && $0.63^{+0.16}_{-0.19}$ && --- && \cite{r-bh-cfm-gou_novamus} \\
XTE~J1652-453 && --- && $< 0.5$ && \cite{r-bh-cfm-1652} \\
XTE~J1550-564 && $0.34 \pm 0.28$ && $0.55^{+0.15}_{-0.22}$ && \cite{r-bh-cfm-xte} \\
LMC~X-3 && $0.25 \pm 0.15$ && --- && \cite{r-bh-cfm-lmcx3} \\
H1743-322 && $0.2 \pm 0.3$ && --- && \cite{r-bh-cfm-h1743} \\
A0620-00 &&  $0.12 \pm 0.19$ && --- && \cite{r-bh-cfm-62} \\
\hspace{0.3cm} XMMU~J004243.6 \hspace{0.3cm} && $< -0.2$ && --- && \cite{r-bh-cfm-m31} \\
\hline 
\end{tabular}}
\vspace{0.4cm}
\caption{Summary of current spin estimates for stellar-mass black holes. The second column refers to the spin estimates via the continuum-fitting method. The third column is for the spin estimates using X-ray reflection spectroscopy (iron line method). See the references in the last column for more details. Note: $^\star$These sources were studied in an early work of the continuum-fitting method, within a more simple model, and therefore the published 1-$\sigma$ error estimates are doubled following~\cite{cfm-5-30-ed}. \label{t-spin-stellar}}
\end{table*}

\begin{table*}[t]
\centering
{\renewcommand{\arraystretch}{1.2}
\begin{tabular}{|ccccc|}
\hline 
Object & \hspace{0.1cm} & $a_*$ (Iron) & \hspace{0.1cm} & \hspace{0.3cm} Principal References \hspace{0.3cm} \\
\hline 
\hspace{0.3cm} IRAS~13224-3809 \hspace{0.3cm} && $> 0.99$ && \cite{r-bh-suzaku} \\
Mrk~110 && $> 0.99$ && \cite{r-bh-suzaku} \\
NGC~4051 && $> 0.99$ && \cite{r-bh-ngc4051} \\
Mrk~509 && $> 0.99$ && \cite{r-bh-mrk509} \\
1H0707-495 && $> 0.98$ && \cite{r-bh-suzaku,r-bh-1h0707} \\
RBS~1124 && $> 0.98$ && \cite{r-bh-suzaku} \\
NGC~3783 && $> 0.98$ && \cite{r-bh-ngc3783} \\
1H0419-577 && $> 0.98$ && \cite{r-bh-suzaku,r-bh-jjc} \\
Fairall~9 && $> 0.97$ && \cite{r-bh-fairall9} \\
NGC~1365 && $0.97^{+0.01}_{-0.04}$ && \cite{r-bh-ngc1365a,r-bh-ngc1365b} \\
Swift~J0501-3239 && $> 0.96$ && \cite{r-bh-suzaku} \\
PDS~456 && $> 0.96$ && \cite{r-bh-suzaku} \\
Ark~564 && $0.96^{+0.01}_{-0.06}$ && \cite{r-bh-suzaku} \\
3C120 && $> 0.95$ && \cite{r-bh-3c120} \\
Mrk~79 && $> 0.95$ && \cite{r-bh-mrk79} \\
NGC~5506 && $0.93^{+0.04}_{-0.04}$ && \cite{r-bh-shangyu} \\
MCG-6-30-15 && $0.91^{+0.06}_{-0.07}$ && \cite{r-bh-mcg63015b} \\
Ton~S180 && $0.91^{+0.02}_{-0.09}$ && \cite{r-bh-suzaku} \\
IRAS~00521-7054 && $> 0.84$ && \cite{r-bh-iras521} \\
Mrk~335 && $0.83^{+0.10}_{-0.13}$ && \cite{r-bh-suzaku,r-bh-mrk335} \\
Ark~120 && $0.81^{+0.10}_{-0.18}$ && \cite{r-bh-suzaku,r-bh-ark120} \\
Swift~J2127+5654 && $0.6^{+0.2}_{-0.2}$ && \cite{r-bh-swift2127} \\
Mrk~841 && $> 0.56$ && \cite{r-bh-suzaku} \\
\hline 
\end{tabular}}
\vspace{0.4cm}
\caption{Summary of current spin estimates for supermassive black holes. See the references in the last column for more details. \label{t-spin-agn}}
\end{table*}

\subsection{X-ray reflection spectroscopy}

Thermal photons from the accretion disk can inverse Compton scatter off free electrons in the so-called corona, which is a generic name to indicate a hotter ($\sim 100$~keV) cloud near the black hole. The corona may be the accretion flow between the inner edge of the disk and the black hole, the base of the jet, or some atmosphere above the accretion disk, but other geometries are also possible. The process produces a power-law spectrum with an exponential cut-off, whose value depends on the coronal temperature. This power-law component illuminates the disk, producing a reflection component. In the rest-frame of the gas in the disk, the reflection spectrum is characterized by narrow fluorescent emission lines in the soft X-ray band, in particular by the iron K$\alpha$ complex at $6.4-7$~keV (depending on the ionization of iron atoms), and by the Compton hump at $20-30$~keV. These fluorescent emission lines are broadened and skewed when observed far from the source as a combination of relativistic effects occurring in the strong gravity region and depending on the exact point of emission in the disk. X-ray reflection spectroscopy refers to the analysis of this reflection component~\cite{brenneman,reynolds}. This technique is also called the iron line method because the iron K$\alpha$ line is usually the most prominent feature in the reflection spectrum, but any spin measurement requires the analysis of the whole reflection spectrum, not only of the iron line. Spin measurements are potentially possible even when the iron K$\alpha$ line is very weak as a result of fully ionized iron atoms.

A summary of current spin measurements of stellar-mass and supermassive black holes using X-ray reflection spectroscopy are reported in Tab.~\ref{t-spin-stellar} (third column) and Tab.~\ref{t-spin-agn}, respectively. It is surely remarkable that the spin measurements of supermassive black holes suggest that most of these objects are rotating very fast, while this is not the case for stellar-mass black holes. This is probably the combination of a few effects:
\begin{enumerate}
\item Fast-rotating black holes are brighter and therefore it is easier to measure their spin. While this is not crucial for stellar-mass black holes, as they are already quite bright sources, it is important for supermassive black holes an introduces and observational bias. 
\item Accretion from disk can spin a black hole up, potentially to the Thorne limit $a_*^{\rm Th} = 0.998$ in the case of a thin disk~\cite{thorne74}. Supermassive black holes can thus have a spin parameter very close to 1, because they may have swallowed a significant amount of mass from their accretion disk. This is not the case for stellar-mass black holes, whose spin parameter is expected to have a value close to that at the time of the formation of the black hole. For objects with a low mass companion star, the spin value cannot increase much even swallowing the whole stellar companion. For objects with a high mass companion star, the latter has a too short lifetime and the black hole cannot swallow enough material to change its spin even accreting at the Eddington limit.  
\item Reflection models are characterized by a large number of parameters that must be inferred from the fit of the spectrum. Only when the value of the spin parameter is very high relativistic effects in the reflection spectrum are strong enough to break parameter degeneracy~\cite{risaliti-sim}. Low spin measurements of supermassive black holes are often obtained by imposing some strong assumptions on the value of other model parameters and they should thus be taken with great caution.
\item Current spin measurements employ reflection models in which the accretion disk is supposed to be infinitesimally thin and with the inner edge at the ISCO radius. However, the thickness of the disk is finite and increases as the mass accretion rate increases. The inner edge of the disk is thought to be at the ISCO radius when the accretion luminosity is between 5\% to 30\% of the Eddington limit~\cite{cfm-5-30-ed}, while we know that most supermassive black holes have higher accretion rates~\cite{brenn13}. When the mass accretion rate is high, the disk becomes geometrically thick and the inner edge can be inside the ISCO radius. Moreover, current reflection models ignore the radiation emitted by the material in the plunging region, between the inner edge of the disk and the black hole. These two effects can lead to overestimate the black hole spin.  
\end{enumerate}

\subsection{Other approaches}

The measurement of the QPO frequencies could potentially provide quite precise measurements of black hole spins, see for instance~\cite{motta13}. However, at present we do not know the exact mechanism responsible for these QPOs and different models provide different measurements, so the technique is not yet mature to measure black hole spins because we do not which model, if any, is the correct one.

Current gravitational wave data of the coalescence of stellar-mass black holes with ground-based laser interferometers already provide estimates of the spin parameter of the final black hole with a precision of order 10\%~\cite{LIGO,LIGO2}. Future observations with space-based laser interferometers promise excellent spin measurements of supermassive black holes at the level of 0.01\%~\cite{EMRIspin1,EMRIspin2}. Similar measurements will be possible from the observations of EMRIs, since an experiment like LISA will be able to observe the signal of an EMRI for $\sim 10^5$ cycles, leading to a very good signal-to-noise ratio.


\section{Testing Fundamental Physics With Black Holes}

\subsection{Motivations}

Einstein's theory of general relativity has been extensively tested in weak gravitational fields~\cite{will}, while the strong gravity regime is still largely unexplored. The spacetime around astrophysical black holes should be well described by the Kerr solution, but deviations from standard predictions can be possible if general relativity is not the correct theory of gravity, in the presence of exotic fields, or in the case of macroscopic quantum gravity effects~\cite{capozziello,herdeiro,giddings}.

Black hole systems are ideal laboratories for testing fundamental physics in strong gravitational fields. This can be done either with electromagnetic or gravitational wave techniques, and the two methods are complementary because they test different sectors of the theory. Electromagnetic tests, strictly speaking, can verify the motion of massive and massless particles in the strong gravitational field around a black hole and check whether it is consistent with the predictions of general relativity. Gravitational wave tests are sensitive to the evolution of the spacetime metric and thus can directly check the predictions of Einstein's equations. For example, a new coupling between the matter and the gravity sectors leading to departure from geodesic motion or variations of fundamental constants may leave an observational signature in the electromagnetic spectrum and not in the gravitational wave one. Modified gravity theories in which uncharged black holes are still described by the Kerr solution may predict the same electromagnetic spectrum as general relativity but have a different gravitational wave signal, because the spacetime metric is the same but the field equations are different.

Let us note it is not true, as instead it is often believed, that tests of general relativity require the study of small effects in the electromagnetic or gravitational wave spectra, difficult to observe and subdominant with respect to the uncertainties of the theoretical models. For example, in the case of electromagnetic tests we can note that reflection spectra from accretion disks around boson stars or wormholes can be qualitatively different from those around Kerr black holes and it is relatively easy to rule out some extreme scenarios~\cite{menglei-16,yueying-16}. Gravitational waves predicted by other theories of gravity can be very different from those expected in general relativity, and even in this case it is possible to obtain very strong constraint on some models~\cite{yunes}.

\subsection{Electromagnetic tests}

In principle, any astrophysical model requiring the Kerr metric to describe some component of the electromagnetic spectrum of an accreting black hole can be extended to test fundamental physics with black holes. From the comparison of the theoretical predictions of the astrophysical model with the data of the source, it is potentially possible to probe new physics by measuring the parameters of the model.

As of now, the most promising electromagnetic technique for testing fundamental physics with black holes is probably X-ray reflection spectroscopy, and surely the only one with quantitative observational constraints already published~\cite{ref1,ref2,ref3,ref4,ref4b,ref5}. Usually, in these tests one assume geodesic motion and compare the theoretical model with observational data to constrain possible deviations from the Kerr metric. However, it is also possible to test the Einstein Equivalence Principle by verifying that atomic physics in the strong gravity region around a black hole is the same as in our laboratories on Earth~\cite{ref6}. Current efforts are devoted to improve the theoretical models to reduce and quantify current systematic uncertainties. It is also very important to select the most suitable sources and data to minimize the systematic uncertainties of the model and maximize relativistic features in the reflection spectrum. For this reason, it is extremely important to select clean sources, without intrinsic absorption, and with the inner edge of the disk as close to possible to the compact object. For more details, see the discussions in Refs.~\cite{ref7,ref8}.

Like X-ray reflection spectroscopy, even the continuum-fitting method can be extended to test the nature of astrophysical black holes~\cite{cfm1,cfm2}. However, this technique can only be used to test stellar-mass black holes, it requires independent measurements of the black hole mass, distance, and inclination angle of the disk (three quantities that are currently difficult to estimate and are often affected by systematic effects), and the thermal component of the disk has a so simple shape that it seems difficult to break the degeneracy among the spin parameter and possible deviations from the Kerr solution~\cite{cfm3}.

The detection of the shadow of the supermassive black hole in M87 with the Event Horizon Telescope has opened the possibility of testing the Kerr metric with accurate detections of black hole shadow~\cite{EHT}. However, the current image can only rule out very exotic scenarios~\cite{sh1,sh2,sh3} and it is not clear the future progress in this direction considering that the inclination angle of the spin of the black hole in M87 with our line of sight is presumably very small and any shadow image will likely be very circular, which makes it impossible to distinguish the Kerr metric from alternative solutions.

Other electromagnetic tests proposed in literature include the study of pulsars orbiting black holes~\cite{pulsar1,pulsar2}, the analysis of QPOs in the spectrum of accreting black holes~\cite{qpo,qpo2}, the detection of blobs of plasma orbiting a black hole~\cite{blob}, the observation of jets powered by the rotational energy of black holes~\cite{jet}, etc.

\subsection{Gravitational wave tests}

Gravitational wave tests are the only method to probe the dynamical strong field regime. Model-independent tests check the consistency of different measurements of the parameters of the system assuming general relativity~\cite{LIGO2}. In model-dependent tests, we compare the theoretical predictions of a theory with observational data to rule out a model or constrain the deviations from general relativity~\cite{yunes}. The disadvantage of model-dependent tests is that there are many modified theories of gravity, so one should repeat the analysis for every model, and the theoretical predictions are difficult to calculate.

Current observations of gravitational waves from black holes are limited to the detection of the coalescence of stellar-mass black holes, where it is possible to study the inspiral and ringdown signals~\cite{berti2,berti3}. Future space-based laser interferometers promise to detect gravitational waves from EMRIs, namely the gravitational waves emitted by a stellar-mass compact object (like a stellar-mass black hole, a neutron star, or a white dwarf) orbiting a supermassive black hole of millions Solar masses. A similar system is very suitable for precision measurements of the spacetime metric around the supermassive black hole because the signal can be observed for years, leading to a high signal-to-noise ratio, and the calculations of the waveforms is relatively easy due to the large difference between the masses of the two bodies~\cite{EMRI}.


\section{Concluding Remarks}

Black holes are among the most amazing objects that can be found in the Universe today and represent ideal laboratories for testing fundamental physics. In the past 50~years, the astronomy community has convinced about the existence of these objects and has started looking for more and more candidates employing different detection techniques. In the past 10-15~years, many efforts have been devoted to measure the spin of these objects and study their host environment assuming standard physics. In the past few years, there have been an increasing number of studies devoted to use astrophysical black holes to test fundamental physics. It is possible that the next decade will be the beginning of a ``Golden Age'' for observational black holes, just like 20~years ago it was the beginning of a golden age for observational cosmology, thanks the advent of new observational facilities and unprecedented high-quality data.


{\bf Acknowledgments --}
This work was supported by the Innovation Program of the Shanghai Municipal Education Commission, Grant No.~2019-01-07-00-07-E00035, and Fudan University, Grant No.~IDH1512060.



\begin{thebibliography}{99}

\bibitem{einstein} 
  A.~Einstein,
  Annalen Phys.\  {\bf 49}, 769 (1916)
  [Annalen Phys.\  {\bf 14}, 517 (2005)].

\bibitem{schwarzschild} 
  K.~Schwarzschild,
  Sitzungsber.\ Preuss.\ Akad.\ Wiss.\ Berlin {\bf 1916}, 189 (1916)
  [physics/9905030].

\bibitem{reissner} 
  H.~Reissner,
  Ann.\ Phys.\  {\bf 59}, 106 (1916).
  
\bibitem{nordstrom} 
  G.~Nordstr\"om,
  Proc.\ Kon.\ Ned.\ Akad.\ Wet.\  {\bf 20}, 1238 (1918).  

\bibitem{finkelstein} 
  D.~Finkelstein,
  Phys.\ Rev.\  {\bf 110}, 965 (1958).
  
\bibitem{kerr} 
  R.~P.~Kerr,
  Phys.\ Rev.\ Lett.\  {\bf 11}, 237 (1963).  
  
\bibitem{newman} 
  E.~T.~Newman, R.~Couch, K.~Chinnapared, A.~Exton, A.~Prakash and R.~Torrence,
  J.\ Math.\ Phys.\  {\bf 6}, 918 (1965).  
  
\bibitem{israel} 
  W.~Israel,
  Phys.\ Rev.\  {\bf 164}, 1776 (1967).
  
\bibitem{carter} 
  B.~Carter,
  Phys.\ Rev.\ Lett.\  {\bf 26}, 331 (1971).

\bibitem{robinson} 
  D.~C.~Robinson,
  Phys.\ Rev.\ Lett.\  {\bf 34}, 905 (1975).  
  
\bibitem{chandrasekhar} 
  S.~Chandrasekhar,
  Astrophys.\ J.\  {\bf 74}, 81 (1931).

\bibitem{oppenheimer} 
  J.~R.~Oppenheimer and G.~M.~Volkoff,
  Phys.\ Rev.\  {\bf 55}, 374 (1939).  
  
\bibitem{zeldovich} 
  Y.~B.~Zeldovich,
  Dokl.\ Akad.\ Nauk\  {\bf 155}, 67 (1964)
  [Sov.\ Phys.\ Dokl. {\bf 9}, 195 (1964)].

\bibitem{salpeter} 
  E.~E.~Salpeter,
  Astrophys.\ J.\  {\bf 140}, 796 (1964).  
  
\bibitem{bowyer} 
  S.~Bowyer, E.~T.~Byram, T.~A.~Chubb and H.~Friedman,
  Science {\bf 147}, 394 (1965).

\bibitem{bolton} 
  C.~T.~Bolton,
  Nature {\bf 235}, 271 (1972).
  
\bibitem{webster} 
  B.~L.~Webster and P.~Murdin,
  Nature {\bf 235}, 37 (1972).    
  
\bibitem{remillard} 
  R.~A.~Remillard and J.~E.~McClintock,
  Ann.\ Rev.\ Astron.\ Astrophys.\  {\bf 44}, 49 (2006)
  [astro-ph/0606352].
  
\bibitem{kormendy} 
  J.~Kormendy and D.~Richstone,
  Ann.\ Rev.\ Astron.\ Astrophys.\  {\bf 33}, 581 (1995).   
  
\bibitem{zhang} 
  S.~N.~Zhang, W.~Cui and W.~Chen,
  Astrophys.\ J.\  {\bf 482}, L155 (1997)
  [astro-ph/9704072].

\bibitem{mcclintock} 
  J.~E.~McClintock, R.~Narayan and J.~F.~Steiner,
  Space Sci.\ Rev.\  {\bf 183}, 295 (2014)
  [arXiv:1303.1583 [astro-ph.HE]].
  
\bibitem{brenneman} 
  L.~W.~Brenneman and C.~S.~Reynolds,
  Astrophys.\ J.\  {\bf 652}, 1028 (2006)
  [astro-ph/0608502].  
  
\bibitem{reynolds} 
  C.~S.~Reynolds,
  Space Sci.\ Rev.\  {\bf 183}, 277 (2014)
  [arXiv:1302.3260 [astro-ph.HE]].  
  
\bibitem{bambi} 
  C.~Bambi,
  Rev.\ Mod.\ Phys.\  {\bf 89}, 025001 (2017)
  [arXiv:1509.03884 [gr-qc]].  

\bibitem{yagi} 
  K.~Yagi and L.~C.~Stein,
  Class.\ Quant.\ Grav.\  {\bf 33}, 054001 (2016)
  [arXiv:1602.02413 [gr-qc]].
  
\bibitem{cardoso} 
  V.~Cardoso and L.~Gualtieri,
  Class.\ Quant.\ Grav.\  {\bf 33}, 174001 (2016)
  [arXiv:1607.03133 [gr-qc]].  

\bibitem{LIGO} 
  B.~P.~Abbott {\it et al.} [LIGO Scientific and Virgo Collaborations],
  Phys.\ Rev.\ Lett.\  {\bf 116}, 061102 (2016)
  [arXiv:1602.03837 [gr-qc]].

\bibitem{EHT} 
  K.~Akiyama {\it et al.} [Event Horizon Telescope Collaboration],
  Astrophys.\ J.\  {\bf 875}, L1 (2019).   

\bibitem{price} 
  R.~H.~Price,
  Phys.\ Rev.\ D {\bf 5}, 2419 (1972).

\bibitem{bambi2} 
  C.~Bambi, D.~Malafarina and N.~Tsukamoto,
  Phys.\ Rev.\ D {\bf 89}, 127302 (2014)
  [arXiv:1406.2181 [gr-qc]].

\bibitem{bambi3} 
  C.~Bambi,
  Annalen Phys.\  {\bf 530}, 1700430 (2018)
  [arXiv:1711.10256 [gr-qc]].
  
\bibitem{bambi4} 
  C.~Bambi, A.~D.~Dolgov and A.~A.~Petrov,
  JCAP {\bf 0909}, 013 (2009)
  [arXiv:0806.3440 [astro-ph]].  
  
\bibitem{ntmod}
  I.~D.~Novikov and K.~S.~Thorne,
  {\it Astrophysics and black holes}, in {\it Black Holes}, edited by C.~De~Witt and B.~De~Witt
  (Gordon and Breach, New York, New York, 1973).    
  
\bibitem{mass-bh1}
  C.~E.~Rhoades and R.~Ruffini,
  Phys.\ Rev.\ Lett.\  {\bf 32}, 324 (1974).   
  
\bibitem{mass-bh2}
  J.~M.~Lattimer,
  Ann.\ Rev.\ Nucl.\ Part.\ Sci.\  {\bf 62}, 485 (2012)
  [arXiv:1305.3510 [nucl-th]].  
 
\bibitem{mass-gap} 
  W.~M.~Farr, N.~Sravan, A.~Cantrell, L.~Kreidberg, C.~D.~Bailyn, I.~Mandel and V.~Kalogera,
  Astrophys.\ J.\  {\bf 741}, 103 (2011)
  [arXiv:1011.1459 [astro-ph.GA]].  
  
\bibitem{m1} 
  K.~Belczynski, T.~Bulik, C.~L.~Fryer, A.~Ruiter, J.~S.~Vink and J.~R.~Hurley,
  Astrophys.\ J.\  {\bf 714}, 1217 (2010)
  [arXiv:0904.2784 [astro-ph.SR]].  

\bibitem{m2} 
  A.~Heger and S.~E.~Woosley,
  Astrophys.\ J.\  {\bf 567}, 532 (2002)
  [astro-ph/0107037].

\bibitem{m3} 
  A.~Heger, C.~L.~Fryer, S.~E.~Woosley, N.~Langer and D.~H.~Hartmann,
  Astrophys.\ J.\  {\bf 591}, 288 (2003)
  [astro-ph/0212469].

\bibitem{m4} 
  M.~Spera, M.~Mapelli and A.~Bressan,
  Mon.\ Not.\ Roy.\ Astron.\ Soc.\  {\bf 451}, no. 4, 4086 (2015)
  [arXiv:1505.05201 [astro-ph.SR]].  
  
\bibitem{bhnum} 
  E.~P.~J.~van den Heuvel,
  {\it Endpoints of stellar evolution: The incidence of stellar mass black holes in the galaxy},
  in ``Environment Observation and Climate Modelling Through International Space Projects'', 29 (1992).  
  
\bibitem{casares} 
  J.~Casares and P.~G.~Jonker,
  Space Sci.\ Rev.\  {\bf 183}, 223 (2014)
  [arXiv:1311.5118 [astro-ph.HE]].  
  
\bibitem{small1} 
  L.~Ferrarese {\it et al.},
  Astrophys.\ J.\  {\bf 644}, L21 (2006)
  [astro-ph/0603840].

\bibitem{small2} 
  E.~Gallo, T.~Treu, J.~Jacob, J.~H.~Woo, P.~Marshall and R.~Antonucci,
  Astrophys.\ J.\  {\bf 680}, 154 (2008)
  [arXiv:0711.2073 [astro-ph]].  
  
\bibitem{mass-bmw1} 
  A.~M.~Ghez, S.~Salim, S.~D.~Hornstein, A.~Tanner, M.~Morris, E.~E.~Becklin and G.~Duchene,
  Astrophys.\ J.\  {\bf 620}, 744 (2005)
  [astro-ph/0306130].  
  
\bibitem{mass-bmw2} 
  A.~Boehle {\it et al.},
  Astrophys.\ J.\  {\bf 830}, 17 (2016)
  [arXiv:1607.05726 [astro-ph.GA]].    
  
\bibitem{maoz} 
  E.~Maoz,
  Astrophys.\ J.\  {\bf 494}, L181 (1998)
  [astro-ph/9710309].  
  
\bibitem{wu15} 
  X.-B.~Wu {\it et al.},
  Nature\  {\bf 518}, 512 (2015).  
 
\bibitem{mv} 
  M.~Volonteri,
  Astron.\ Astrophys.\ Rev.\  {\bf 18}, 279 (2010)
  [arXiv:1003.4404 [astro-ph.CO]].  
  
\bibitem{supere} 
  P.~Madau, F.~Haardt and M.~Dotti,
  Astrophys.\ J.\  {\bf 784}, L38 (2014)
  [arXiv:1402.6995 [astro-ph.CO]].  
  
\bibitem{miller03} 
  M.~C.~Miller and E.~J.~M.~Colbert,
  Int.\ J.\ Mod.\ Phys.\ D {\bf 13}, 1 (2004)
  [astro-ph/0308402].  
  
\bibitem{ulxs} 
  E.~J.~M.~Colbert and R.~F.~Mushotzky,
  Astrophys.\ J.\  {\bf 519}, 89 (1999)
  [astro-ph/9901023].
  
\bibitem{supere2} 
  M.~Bachetti {\it et al.},
  Nature {\bf 514}, 202
  [arXiv:1410.3590 [astro-ph.HE]].
  
\bibitem{supere3} 
  K.~Ohsuga and S.~Mineshige,
  Astrophys.\ J.\  {\bf 736}, 2 (2011)
  [arXiv:1105.5474 [astro-ph.HE]].  

\bibitem{interqpo} 
  D.~R.~Pasham, T.~E.~Strohmayer and R.~F.~Mushotzky,
  Nature {\bf 513}, 74 (2014)
  [arXiv:1501.03180 [astro-ph.HE]].  
  
\bibitem{gclu1} 
  K.~Gebhardt, R.~M.~Rich and L.~C.~Ho,
  Astrophys.\ J.\  {\bf 578}, L41 (2002)
  [astro-ph/0209313].  
  
\bibitem{gclu2} 
  K.~Gebhardt, R.~M.~Rich and L.~C.~Ho,
  Astrophys.\ J.\  {\bf 634}, 1093 (2005)
  [astro-ph/0508251].
  
\bibitem{EMRIspin1} 
  L.~Barack and C.~Cutler,
  Phys.\ Rev.\ D {\bf 69}, 082005 (2004)
  [gr-qc/0310125].

\bibitem{EMRIspin2} 
  S.~Babak {\it et al.},
  Phys.\ Rev.\ D {\bf 95}, 103012 (2017)
  [arXiv:1703.09722 [gr-qc]].    
  
\bibitem{kerrbb} 
  L.~X.~Li, E.~R.~Zimmerman, R.~Narayan and J.~E.~McClintock,
  Astrophys.\ J.\ Suppl.\  {\bf 157}, 335 (2005)
  [astro-ph/0411583].  
  
\bibitem{cfm-agn1} 
  C.~Done, C.~Jin, M.~Middleton and M.~Ward,
  Mon.\ Not.\ Roy.\ Astron.\ Soc.\  {\bf 434}, 1955 (2013)
  [arXiv:1306.4786 [astro-ph.HE]]. 
  
\bibitem{cfm-agn2} 
  D.~M.~Capellupo, G.~Wafflard-Fernandez and D.~Haggard,
  Astrophys.\ J.\  {\bf 836}, L8 (2017)
  [arXiv:1701.07887 [astro-ph.GA]].   
  
\bibitem{cfm-5-30-ed} 
  J.~E.~McClintock, R.~Narayan and J.~F.~Steiner,
  Space Sci.\ Rev.\  {\bf 183}, 295 (2014)
  [arXiv:1303.1583 [astro-ph.HE]].  
  
\bibitem{r-bh-cfm-1915} 
  J.~E.~McClintock, R.~Shafee, R.~Narayan, R.~A.~Remillard, S.~W.~Davis and L.~X.~Li,
  Astrophys.\ J.\  {\bf 652}, 518 (2006)
  [astro-ph/0606076]. 

\bibitem{r-bh-cfm-1915b} 
  J.~M.~Miller {\it et al.},
  Astrophys.\ J.\  {\bf 775}, L45 (2013)
  [arXiv:1308.4669 [astro-ph.HE]].

\bibitem{r-bh-cfm-cyg1} 
  L.~Gou {\it et al.},
  Astrophys.\ J.\  {\bf 742}, 85 (2011)
  [arXiv:1106.3690 [astro-ph.HE]].  
  
\bibitem{r-bh-cfm-cyg2}
  L.~Gou {\it et al.},
  Astrophys.\ J.\  {\bf 790}, 29 (2014)
  [arXiv:1308.4760 [astro-ph.HE]].

\bibitem{r-bh-cfm-cyg3}  
  A.~C.~Fabian {\it et al.},
  Mon.\ Not.\ Roy.\ Astron.\ Soc.\  {\bf 424}, 217 (2012)
  [arXiv:1204.5854 [astro-ph.HE]].   

\bibitem{r-bh-cfm-cyg4} 
  J.~A.~Tomsick {\it et al.},
  Astrophys.\ J.\  {\bf 780}, 78 (2014)
  [arXiv:1310.3830 [astro-ph.HE]]. 

\bibitem{r-bh-cfm-cyg5} 
  M.~L.~Parker {\it et al.},
  Astrophys.\ J.\  {\bf 808}, 9 (2015)
  [arXiv:1506.00007 [astro-ph.HE]].

\bibitem{r-bh-cfm-cyg6} 
  D.~J.~Walton {\it et al.},
  Astrophys.\ J.\  {\bf 826}, 87 (2016)
  [arXiv:1605.03966 [astro-ph.HE]].   

\bibitem{r-bh-cfm-gs1354} 
  A.~M.~El-Batal {\it et al.},
  Astrophys.\ J.\  {\bf 826}, L12 (2016)
  [arXiv:1607.00343 [astro-ph.HE]].  

\bibitem{r-bh-iron-maxi15a} 
  Y.~Xu {\it et al.},
  Astrophys.\ J.\  {\bf 852}, L34 (2018)
  [arXiv:1711.01346 [astro-ph.HE]].

\bibitem{r-bh-iron-maxi15b} 
  J.~M.~Miller {\it et al.},
  Astrophys.\ J.\  {\bf 860}, L28 (2018)
  [arXiv:1806.04115 [astro-ph.HE]].   

\bibitem{r-bh-iron-swift-xu} 
  Y.~Xu {\it et al.},
  arXiv:1805.07705 [astro-ph.HE].    

\bibitem{r-bh-cfm-lmcx1} 
  L.~Gou, J.~E.~McClintock, J.~Liu, R.~Narayan, J.~F.~Steiner, R.~A.~Remillard, J.~A.~Orosz and S.~W.~Davis,
  Astrophys.\ J.\  {\bf 701}, 1076 (2009)
  [arXiv:0901.0920 [astro-ph.HE]].   

\bibitem{r-bh-cfm-lmcx1b}   
  J.~F.~Steiner {\it et al.},
  Mon.\ Not.\ Roy.\ Astron.\ Soc.\  {\bf 427}, 2552 (2012)
  [arXiv:1209.3269 [astro-ph.HE]]. 

\bibitem{r-bh-cfm-gx339} 
  M.~Kolehmainen and C.~Done,
  Mon.\ Not.\ Roy.\ Astron.\ Soc.\  {\bf 406}, 2206 (2010)
  [arXiv:0911.3281 [astro-ph.HE]]. 

\bibitem{r-bh-cfm-gx339b} 
  R.~C.~Reis, A.~C.~Fabian, R.~Ross, G.~Miniutti, J.~M.~Miller and C.~Reynolds,
  Mon.\ Not.\ Roy.\ Astron.\ Soc.\  {\bf 387}, 1489 (2008)
  [arXiv:0804.0238 [astro-ph]].  

\bibitem{r-bh-cfm-gx339c} 
  J.~Garcia {\it et al.},
  Astrophys.\ J.\ {\bf 813}, 84 (2015)
  [arXiv:1505.03607[astro-ph.HE]].   

\bibitem{r-bh-cfm-gx339d} 
  M.~L.~Parker {\it et al.},
  Astrophys.\ J.\  {\bf 821}, L6 (2016)
  [arXiv:1603.03777 [astro-ph.HE]].    

\bibitem{r-bh-cfm-1752} 
  R.~C.~Reis {\it et al.},
  Mon.\ Not.\ Roy.\ Astron.\ Soc.\  {\bf 410}, 2497 (2011)
  [arXiv:1009.1154 [astro-ph.HE]].     

\bibitem{r-bh-cfm-maxi} 
  R.~C.~Reis, J.~M.~Miller, M.~T.~Reynolds, A.~C.~Fabian and D.~J.~Walton,
  Astrophys.\ J.\  {\bf 751}, 34 (2012)
  [arXiv:1111.6665 [astro-ph.HE]].   

\bibitem{r-bh-cfm-liu08} 
  J.~Liu, J.~McClintock, R.~Narayan, S.~Davis and J.~Orosz,
  Astrophys.\ J.\  {\bf 679}, L37 (2008) [Erratum: Astrophys.\ J.\  {\bf 719}, L109 (2010)]
  [arXiv:0803.1834 [astro-ph]].   

\bibitem{r-bh-cfm-sh06} 
  R.~Shafee, J.~E.~McClintock, R.~Narayan, S.~W.~Davis, L.~X.~Li and R.~A.~Remillard,
  Astrophys.\ J.\  {\bf 636}, L113 (2006)
  [astro-ph/0508302].   

 \bibitem{r-bh-cfm-st16}
  J.~F.~Steiner, D.~J.~Walton, J.~A.~Garcia, J.~E.~McClintock, S.~G.~T.~Laycock, M.~J.~Middleton, R.~Barnard and K.~K.~Madsen,
  arXiv:1512.03414 [astro-ph.HE]. 

\bibitem{r-bh-cfm-swift}   
  R.~C.~Reis, A.~C.~Fabian, R.~R.~Ross and J.~M.~Miller,
  Mon.\ Not.\ Roy.\ Astron.\ Soc.\  {\bf 395}, 1257 (2009).
  
\bibitem{r-bh-cfm-1650} 
  D.~J.~Walton, R.~C.~Reis, E.~M.~Cackett, A.~C.~Fabian and J.~M.~Miller,
  Mon.\ Not.\ Roy.\ Astron.\ Soc.\  {\bf 422}, 2510 (2012)
  [arXiv:1202.5193 [astro-ph.HE]].  

\bibitem{r-bh-cfm-gou_novamus}
  Z.~Chen, L.~Gou, J.~E.~McClintock, J.~F.~Steiner, J.~Wu, W.~Xu, J.~Orosz and Y.~Xiang,
  arXiv:1601.00615 [astro-ph.HE]. 

\bibitem{r-bh-cfm-1652} 
  C.~Y.~Chiang, R.~C.~Reis, D.~J.~Walton and A.~C.~Fabian,
  Mon.\ Not.\ Roy.\ Astron.\ Soc.\  {\bf 425}, 2436 (2012)
  [arXiv:1207.0682 [astro-ph.HE]].   

\bibitem{r-bh-cfm-xte} 
  J.~F.~Steiner {\it et al.},
  Mon.\ Not.\ Roy.\ Astron.\ Soc.\  {\bf 416}, 941 (2011)
  [arXiv:1010.1013 [astro-ph.HE]]. 

\bibitem{r-bh-cfm-lmcx3} 
  J.~F.~Steiner, J.~E.~McClintock, J.~A.~Orosz, R.~A.~Remillard, C.~D.~Bailyn, M.~Kolehmainen and O.~Straub,
  Astrophys.\ J.\  {\bf 793}, L29 (2014)
  [arXiv:1402.0148 [astro-ph.HE]].  

\bibitem{r-bh-cfm-h1743} 
  J.~F.~Steiner, J.~E.~McClintock and M.~J.~Reid,
  Astrophys.\ J.\  {\bf 745}, L7 (2012)
  [arXiv:1111.2388 [astro-ph.HE]].   

\bibitem{r-bh-cfm-62} 
  L.~Gou, J.~E.~McClintock, J.~F.~Steiner, R.~Narayan, A.~G.~Cantrell, C.~D.~Bailyn and J.~A.~Orosz,
  Astrophys.\ J.\  {\bf 718}, L122 (2010)
  [arXiv:1002.2211 [astro-ph.HE]].    

\bibitem{r-bh-cfm-m31} 
  M.~Middleton, J.~Miller-Jones and R.~Fender,
  Mon.\ Not.\ Roy.\ Astron.\ Soc.\  {\bf 439}, 1740 (2014)
  [arXiv:1401.1829 [astro-ph.HE]].  
  
\bibitem{r-bh-suzaku} 
  D.~J.~Walton, E.~Nardini, A.~C.~Fabian, L.~C.~Gallo and R.~C.~Reis,
  Mon.\ Not.\ Roy.\ Astron.\ Soc.\  {\bf 428}, 2901 (2013)
  [arXiv:1210.4593 [astro-ph.HE]].  

\bibitem{r-bh-ngc4051} 
  A.~R.~Patrick, J.~N.~Reeves, D.~Porquet, A.~G.~Markowitz, V.~Braito and A.~P.~Lobban,
  Mon.\ Not.\ Roy.\ Astron.\ Soc.\  {\bf 426}, 2522 (2012)
  [arXiv:1208.1150 [astro-ph.HE]].   

\bibitem{r-bh-mrk509} 
  J.~A.~Garcia {\it et al.},
  Astrophys.\ J.\  {\bf 871}, 88 (2019)
  [arXiv:1812.03194 [astro-ph.HE]].

\bibitem{r-bh-1h0707} 
  A.~Zoghbi, A.~Fabian, P.~Uttley, G.~Miniutti, L.~Gallo, C.~Reynolds, J.~Miller and G.~Ponti,
  Mon.\ Not.\ Roy.\ Astron.\ Soc.\  {\bf 401}, 2419 (2010)
  [arXiv:0910.0367 [astro-ph.HE]].    

\bibitem{r-bh-ngc3783} 
  L.~W.~Brenneman {\it et al.},
  Astrophys.\ J.\  {\bf 736}, 103 (2011)
  [arXiv:1104.1172 [astro-ph.HE]]. 
  
\bibitem{r-bh-jjc} 
  J.~Jiang, D.~J.~Walton, A.~C.~Fabian and M.~L.~Parker,
  Mon.\ Not.\ Roy.\ Astron.\ Soc.\  {\bf 483}, 2958 (2019)
  [arXiv:1811.10932 [astro-ph.HE]].    

\bibitem{r-bh-fairall9} 
  A.~Lohfink {\it et al.},
  Astrophys.\ J.\  {\bf 821}, 11 (2016)
  [arXiv:1602.05589 [astro-ph.GA]].

\bibitem{r-bh-ngc1365a} 
  G.~Risaliti {\it et al.},
  Nature {\bf 494}, 449 (2013)
  [arXiv:1302.7002 [astro-ph.HE]].  

\bibitem{r-bh-ngc1365b} 
  L.~W.~Brenneman, G.~Risaliti, M.~Elvis and E.~Nardini,
  Mon.\ Not.\ Roy.\ Astron.\ Soc.\  {\bf 429}, 2662 (2013)
  [arXiv:1212.0772 [astro-ph.HE]].  

\bibitem{r-bh-3c120} 
  A.~M.~Lohfink {\it et al.},
  Astrophys.\ J.\  {\bf 772}, 83 (2013)
  [arXiv:1305.4937 [astro-ph.HE]]. 

\bibitem{r-bh-mrk79} 
  L.~C.~Gallo, G.~Miniutti, J.~M.~Miller, L.~W.~Brenneman, A.~C.~Fabian, M.~Guainazzi and C.~S.~Reynolds,
  Mon.\ Not.\ Roy.\ Astron.\ Soc.\  {\bf 411}, 607 (2011)
  [arXiv:1009.2987 [astro-ph.HE]].    

\bibitem{r-bh-shangyu} 
  S.~Sun, M.~Guainazzi, Q.~Ni, J.~Wang, C.~Qian, F.~Shi, Y.~Wang and C.~Bambi,
  Mon.\ Not.\ Roy.\ Astron.\ Soc.\  {\bf 478}, 1900 (2018)
  [arXiv:1704.03716 [astro-ph.HE]].  

\bibitem{r-bh-mcg63015b} 
  A.~Marinucci {\it et al.},
  Astrophys.\ J.\  {\bf 787}, 83 (2014)
  [arXiv:1404.3561 [astro-ph.HE]].  

\bibitem{r-bh-iras521} 
  Y.~Tan, J.~Wang, X.~Shu and Y.~Zhou,
  Astrophys.\ J.\  {\bf 747}, L11 (2012)
  [arXiv:1202.0400 [astro-ph.HE]].  

\bibitem{r-bh-mrk335} 
  M.~L.~Parker {\it et al.},
  Mon.\ Not.\ Roy.\ Astron.\ Soc.\  {\bf 443}, no. 2, 1723 (2014)
  [arXiv:1407.8223 [astro-ph.HE]].  

\bibitem{r-bh-ark120} 
  E.~Nardini, A.~C.~Fabian, R.~C.~Reis and D.~J.~Walton,
  Mon.\ Not.\ Roy.\ Astron.\ Soc.\  {\bf 410}, 1251 (2011)
  [arXiv:1008.2157 [astro-ph.HE]].

\bibitem{r-bh-swift2127} 
  G.~Miniutti, F.~Panessa, A.~De Rosa, A.~C.~Fabian, A.~Malizia, M.~Molina, J.~M.~Miller and S.~Vaughan,
  Mon.\ Not.\ Roy.\ Astron.\ Soc.\  {\bf 398}, 255 (2009)
  [arXiv:0905.2891 [astro-ph.HE]].  

\bibitem{thorne74} 
  K.~S.~Thorne,
  Astrophys.\ J.\  {\bf 191}, 507 (1974).  
  
\bibitem{risaliti-sim} 
  E.~S.~Kammoun, E.~Nardini and G.~Risaliti,
  Astron.\ Astrophys.\  {\bf 614}, A44 (2018)
  [arXiv:1802.06800 [astro-ph.HE]].  
  
\bibitem{brenn13} 
  L.~Brenneman,
  {\it Measuring Supermassive Black Hole Spins in Active Galactic Nuclei} (Springer, New York, 2013),
  doi:10.1007/978-1-4614-7771-6
  [arXiv:1309.6334 [astro-ph.HE]].  
  
\bibitem{motta13} 
  S.~E.~Motta, T.~M.~Belloni, L.~Stella, T.~Muñoz-Darias and R.~Fender,
  Mon.\ Not.\ Roy.\ Astron.\ Soc.\  {\bf 437}, 2554 (2014)
  [arXiv:1309.3652 [astro-ph.HE]].  
  
\bibitem{LIGO2} 
  B.~P.~Abbott {\it et al.} [LIGO Scientific and Virgo Collaborations],
  Phys.\ Rev.\ Lett.\  {\bf 116}, 221101 (2016)
  [Erratum: Phys.\ Rev.\ Lett.\  {\bf 121}, 129902 (2018)]
  [arXiv:1602.03841 [gr-qc]].  
     
\bibitem{will} 
  C.~M.~Will,
  Living Rev.\ Rel.\  {\bf 17}, 4 (2014)
  [arXiv:1403.7377 [gr-qc]].

\bibitem{capozziello} 
  S.~Capozziello and M.~De Laurentis,
  Phys.\ Rept.\  {\bf 509}, 167 (2011)
  [arXiv:1108.6266 [gr-qc]].

\bibitem{herdeiro} 
  C.~A.~R.~Herdeiro and E.~Radu,
  Phys.\ Rev.\ Lett.\  {\bf 112}, 221101 (2014)
  [arXiv:1403.2757 [gr-qc]].

\bibitem{giddings} 
  S.~B.~Giddings,
  Nature Astronomy {\bf 1}, 0067 (2017)
  [arXiv:1703.03387 [gr-qc]].  
  
\bibitem{menglei-16} 
  M.~Zhou, A.~Cardenas-Avendano, C.~Bambi, B.~Kleihaus and J.~Kunz,
  Phys.\ Rev.\ D {\bf 94}, 024036 (2016)
  [arXiv:1603.07448 [gr-qc]].

\bibitem{yueying-16} 
  Y.~Ni, M.~Zhou, A.~Cardenas-Avendano, C.~Bambi, C.~A.~R.~Herdeiro and E.~Radu,
  JCAP {\bf 1607}, 049 (2016)
  [arXiv:1606.04654 [gr-qc]].  
  
\bibitem{yunes} 
  N.~Yunes, K.~Yagi and F.~Pretorius,
  Phys.\ Rev.\ D {\bf 94}, 084002 (2016)
  [arXiv:1603.08955 [gr-qc]].  
  
\bibitem{ref1} 
  C.~Bambi, A.~Cardenas-Avendano, T.~Dauser, J.~A.~Garcia and S.~Nampalliwar,
  Astrophys.\ J.\  {\bf 842}, 76 (2017)
  [arXiv:1607.00596 [gr-qc]].

\bibitem{ref2} 
  Z.~Cao, S.~Nampalliwar, C.~Bambi, T.~Dauser and J.~A.~Garcia,
  Phys.\ Rev.\ Lett.\  {\bf 120}, 051101 (2018)
  [arXiv:1709.00219 [gr-qc]].

\bibitem{ref3} 
  A.~Tripathi {\it et al.},
  Astrophys.\ J.\  {\bf 875}, 56 (2019)
  [arXiv:1811.08148 [gr-qc]].
  
\bibitem{ref4} 
  A.~Tripathi {\it et al.},
  Astrophys.\ J.\  {\bf 874}, 135 (2019)
  [arXiv:1901.03064 [gr-qc]].  
  
\bibitem{ref4b} 
  A.~Tripathi, A.~B.~Abdikamalov, D.~Ayzenberg, C.~Bambi and S.~Nampalliwar,
  Phys.\ Rev.\ D {\bf 99}, 083001 (2019)
  [arXiv:1903.04071 [gr-qc]].  
  
\bibitem{ref5} 
  A.~B.~Abdikamalov, D.~Ayzenberg, C.~Bambi, T.~Dauser, J.~A.~Garcia and S.~Nampalliwar,
  Astrophys.\ J.\  (in press)
  arXiv:1902.09665 [gr-qc].   
  
\bibitem{ref6} 
  C.~Bambi,
  JCAP {\bf 1403}, 034 (2014)
  [arXiv:1308.2470 [gr-qc]].  
  
\bibitem{ref7} 
  H.~Liu, A.~B.~Abdikamalov, D.~Ayzenberg, C.~Bambi, T.~Dauser, J.~A.~Garcia and S.~Nampalliwar,
  Phys.\ Rev.\ D (in press)
  [arXiv:1904.08027 [gr-qc]].
  
\bibitem{ref8} 
  A.~B.~Abdikamalov {\it et al.},
  arXiv:1905.08012 [gr-qc].  
  
\bibitem{cfm1} 
  C.~Bambi and E.~Barausse,
  Astrophys.\ J.\  {\bf 731}, 121 (2011)
  [arXiv:1012.2007 [gr-qc]].
  
\bibitem{cfm2} 
  M.~Zhou, A.~B.~Abdikamalov, D.~Ayzenberg, C.~Bambi, H.~Liu and S.~Nampalliwar,
  Phys.\ Rev.\ D {\bf 99}, 104031 (2019)
  [arXiv:1903.09782 [gr-qc]].  
  
\bibitem{cfm3} 
  L.~Kong, Z.~Li and C.~Bambi,
  Astrophys.\ J.\  {\bf 797}, 78 (2014)
  [arXiv:1405.1508 [gr-qc]].    
  
\bibitem{sh1} 
  C.~Bambi and K.~Freese,
  Phys.\ Rev.\ D {\bf 79}, 043002 (2009)
  [arXiv:0812.1328 [astro-ph]].

\bibitem{sh2} 
  C.~Bambi,
  Phys.\ Rev.\ D {\bf 87}, 107501 (2013)
  [arXiv:1304.5691 [gr-qc]].

\bibitem{sh3} 
  C.~Bambi, K.~Freese, S.~Vagnozzi and L.~Visinelli,
  arXiv:1904.12983 [gr-qc].  
  
\bibitem{pulsar1} 
  N.~Wex and S.~Kopeikin,
  Astrophys.\ J.\  {\bf 514}, 388 (1999)
  [astro-ph/9811052].
  
\bibitem{pulsar2} 
  K.~Liu, N.~Wex, M.~Kramer, J.~M.~Cordes and T.~J.~W.~Lazio,
  Astrophys.\ J.\  {\bf 747}, 1 (2012)
  [arXiv:1112.2151 [astro-ph.HE]].
  
\bibitem{qpo} 
  C.~Bambi,
  JCAP {\bf 1209}, 014 (2012)
  [arXiv:1205.6348 [gr-qc]].
  
\bibitem{qpo2} 
  C.~Bambi,
  Eur.\ Phys.\ J.\ C {\bf 75}, 162 (2015)
  [arXiv:1312.2228 [gr-qc]].  
  
\bibitem{blob} 
  Z.~Li and C.~Bambi,
  Phys.\ Rev.\ D {\bf 90}, 024071 (2014)
  [arXiv:1405.1883 [gr-qc]].  
  
\bibitem{jet} 
  C.~Bambi,
  Phys.\ Rev.\ D {\bf 86}, 123013 (2012)
  [arXiv:1204.6395 [gr-qc]]. 
    
\bibitem{berti2} 
  E.~Berti, K.~Yagi and N.~Yunes,
  Gen.\ Rel.\ Grav.\  {\bf 50}, 46 (2018)
  [arXiv:1801.03208 [gr-qc]].  
  
\bibitem{berti3} 
  E.~Berti, K.~Yagi, H.~Yang and N.~Yunes,
  Gen.\ Rel.\ Grav.\  {\bf 50}, 49 (2018)
  [arXiv:1801.03587 [gr-qc]].   
  
\bibitem{EMRI} 
  L.~Barack and C.~Cutler,
  Phys.\ Rev.\ D {\bf 75}, 042003 (2007)
  [gr-qc/0612029].  
  
\end{thebibliography}
\end{document}